# SOLARIS: Solar sail investigation of the Sun

Thierry Appourchaux[1] and Frédéric Auchère[2]
*Institut d'Astrophysique Spatiale, 91405 Orsay Cedex, France*

Ester Antonucci[3]
*Istituto Nazionale di Astrofisica, Osservatorio Astrofisico di Torino, Italy*

Laurent Gizon[4]
*Max-Planck-Institut für Sonnensystemforschung, 37191 Katlenburg-Lindau, Germany and
Institut für Astrophysik, Georg-August-Universität Göttingen, 37077 Göttingen, Germany*

and
Malcolm MacDonald[5]
*University of Strathclyde, Glasgow G4 0LT, United Kingdom*

and
Hirohisa Hara[6] and Takashi Sekii[7]
*National Astronomical Observatory of Japan, Tokyo, Japan*

and
Daniel Moses[8] and Angelos Vourlidas[9]
*Naval Research Laboratory, Washington DC, USA*

**In this paper, we detail the scientific objectives and outline a strawman payload of the SOLAR sail Investigation of the Sun (SOLARIS). The science objectives are to study the 3D structure of the solar magnetic and velocity field, the variation of total solar irradiance with latitude, and the structure of the corona. We show how we can meet these science objective using solar-sail technologies currently under development. We provide a tentative mission profile considering several trade-off approaches. We also provide a tentative mass budget breakdown and a perspective for a programmatic implementation.**

## I. Introduction

Understanding the origins of solar magnetic activity has been at the forefront of solar and stellar physics since the discovery of the 11-year sunspot cycle nearly two centuries ago. Unraveling this mystery has broad

---

[1] EUI (Co-PI), PHI (Lead Co-I), SPICE (Co-I) (All Solar orbiter), Deputy Director, Solar and Stellar Physics Department, Bâtiment 121
[2] EUI (Solar Orbiter) Instrument Scientist, Solar and Stellar Physics Department, Bâtiment 121
[3] METIS (Solar Orbiter) Principal Investigator
[4] Director, Department Solar and Stellar Interiors
[5] Senior Lecturer, Dept. of Mechanical and Aerospace Engineering, Lord Hope Bld, 141 St James Road.
[6] Associate Professor
[7] Associate Professor, Solar-C project office
[8] Solar & Heliospheric Physics Branch, Space Science Division
[9] Head, Coronal & Heliophysics Section, Solar & Heliospheric Physics Branch, Space Science Division

implications not only for promoting a deeper knowledge of the Sun itself but also for understanding the Sun's influence on the heliosphere, the geospace environment, and potentially the Earth's climate system. Such influences regulate space weather, with increasing economic impacts on our technological society as our reliance on telecommunications systems, power grids, and airline travel continues to grow. As a readily observable example of an astrophysical magneto-hydrodynamic (MHD) dynamo, the Sun offers unique insights into the generation of magnetic fields by turbulent plasma flows throughout the universe, from planetary and stellar interiors to stellar and galactic accretion disks to interstellar clouds.

The global magnetic polarity of the Sun reverses during each 11-year sunspot cycle so that the overall period of the solar magnetic activity cycle is 22 years. It is a formidable challenge to understand how such remarkable regularity arises from the highly turbulent conditions of the solar convection zone and how magnetic flux emerges from the solar interior to energize the solar atmosphere and to drive solar variability. Large-scale flows (differential rotation and meridional circulations) established by turbulent convection, plasma instabilities, and nonlinear feedbacks all play an important role, spanning many orders of magnitude in spatial and temporal scales.

Modern solar observations coupled with sophisticated theoretical and numerical models have yielded important insights into many aspects of solar magnetism but the basic physical mechanisms responsible for generating these fields are still not understood. To achieve scientific progress in our understanding the Sun and the fundamental problems of cosmic magnetism we need to redouble both theoretical and observational efforts. On the observational side, measuring solar internal flows is of the greatest importance. For this task helioseismology has proven to be a powerful tool.

Helioseismic measurements are based on surface wavefield data, normally and preferably temporal series of photospheric Dopplergrams, which are then analyzed to probe the solar interior structure and flows. With the so-called global methods, the wavefield data are used to measure (mainly acoustic) eigenfrequencies of the Sun. The eigenfrequencies are then analyzed, often by way of inverse methods, to probe the solar interior for thermal and dynamical structure of high degrees of symmetry, such as the spherically symmetric distribution of sound speed, or differential rotation as the axisymmetric component of flows. With new local methods, wavefield data are used to measure local resonant properties or wave propagation time for a given pair of points, by cross-correlating local wavefields. These travel-time data are then analyzed to probe the interior for local and/or asymmetric structures, such as meridional flow, convection and flows around active regions [5].

Differential rotation and meridional flows have already been measured by such helioseismology techniques, up to about 60° latitude with a typical uncertainty of the order of a few m s$^{-1}$ [10]. It is essential to extend this measurement to the polar region, partly because these measurements will validate our understanding of dynamics of the Sun as a whole, and partly because the polar region is where the magnetic flux reverses and the meridional flow, which plays an important role in carrying the magnetic flux, should turn in towards the solar interior.

Another related mystery is the total solar irradiance variation over the solar cycle. The total irradiance of other solar-like stars with similar activity level, varies around 0.3 per cent, on average, over their activity cycles. On the other hand, the solar irradiance varies only by 0.1 per cent. There is a well-founded suspicion that the solar irradiance depends on latitude, thereby creating a great interest in measuring the solar irradiance from high heliographic latitudes.

Because of the orbit characteristics of the vast majority of spacecraft, the solar flux has predominantly been measured at Earth or generally in the plane of the ecliptic. Therefore, the existing data cannot directly demonstrate whether the latitudinal distribution of the extreme-ultraviolet (EUV) solar flux is anisotropic. Indeed, the non-uniform distribution of very contrasted EUV bright features (i.e., active regions) and dark features (i.e., coronal holes) at the surface of the Sun produces both the obvious rotational (or longitudinal) modulation of the flux and also a strong latitudinal anisotropy. This has been demonstrated by [4], but no direct measurements exist. Although largely ignored up to now, the latitudinal anisotropy has several important implications. A first consequence is that the total EUV output of the Sun is currently overestimated, which affects comparisons with other stars. The EUV latitudinal anisotropies also affect the photo-ionization rates of helium that are used do derive the helium abundance in the local interstellar medium (LISM) from measurements made in-situ in the heliosphere (e.g. by GAS/Ulysses). The helium abundance in the LISM is a major tracer of galactic evolution. Measuring the solar flux in at least two EUV wavebands will enable to test and to complete the models of the irradiance anisotropies derived from in-Ecliptic observations.

Many facts point to a weak coupling between the magnetic fields in the northern and southern hemispheres of the Sun. The reversals of the magnetic field polarity at the poles do not occur simultaneously, with one reversal delayed by as much as two years relative to the other. The Emergence of the magnetic field in the photosphere was organized in a persistent large-scale pattern with different rotation periods in the northern and southern hemispheres, in cycles 20 and 21, pointing to a weak interdependence of the field systems originating in the two hemispheres [1]. Galactic cosmic rays modulated by solar activity are affected by a north-south gradient that reverses at the reversal of the solar magnetic dipole, indicating a different level of activity, and thus magnetic flux, in the two hemispheres. A significant north-south asymmetry in the EUV solar flux was discovered in solar cycle 23. Measurements of solar irradiance at the poles have not yet been possible and therefore we cannot assess whether this quantity is the same at the north and the south pole. Thus observations suggest that the magnetic field systems generated in the two hemispheres can indeed evolve independently, although they tend to exchange characteristics at the polarity reversal, following the magnetic cycle.

Recent results show that the corona itself is site of a vast variety of fluctuations. The coronal images, obtained with STEREO, show the existence of quasi-periodic non-stationary density variations characterized by a wide range of temporal and spatial scales and strongly confined by the magnetic topology. In closed field line structures, at mid and low latitudes, the density variations might be interpreted as due to slow standing magneto-acoustic waves excited by the convective super-granular motions [12]. In the outer corona, observed in the ultraviolet line HI Ly $\alpha$ with UVCS-SOHO, density fluctuations are coherent and persistent in the slow coronal wind, at mid and low latitudes, whilst in the fast wind, at high latitudes, they are stochastic and non-correlated [11]. The analysis and interpretation of this kind of coronal phenomena is still in its infancy and a complete scenario which links the coronal fluctuations back to their magnetic roots in the photosphere and which helps to assess their role in the energetics of the solar corona and the solar wind, is still to be developed.

Finally, several aspects of the large-scale magnetic configuration of the corona and the interplay between explosive activity and background magnetic field remain either elusive or ambiguous. Questions such as: *what is the 3D structure of the global corona? What is the longitudinal extent of coronal mass ejections (CMEs)? What is the magnetic connectivity between the outgoing CME and low corona? Are post-CME rays the plasma envelopes of the post-CME current sheet predicted by theories of eruptions?* remain open issues despite several years of multi-viewpoint observations from STEREO. For example, both STEREO spacecraft lie on the ecliptic and hence can provide information for the density distribution of the background corona and CMEs only normal to the ecliptic. For this reason, inversion techniques cannot be used reliably unless out-the ecliptic information on the longitudinal spread of coronal structures is available. Such observations provided by a wide angle EUV imager will probe directly the longitudinal evolution of eruptions from the very low corona (< 70,000 km) to the outer corona (~15 $R_s$) and verify whether there exists a super-radial expansion phase during the early part of the event, as postulated by [9], whether the changes in the CME angular width are due simply to projection effects (i.e. rotation) or due to interaction with the ambient coronal structures, whether CMEs can be deflected longitudinally by nearby coronal holes [8], and whether CMEs can have angular momentum coupling with the Sun as proposed by [7]. All these effects can have important consequences for understanding the development of CMEs and eventually predicting their geoeffectiveness. Wide-angle EUV observations of post-CME structures, from high latitudes, will finally uncover the natures of these features. They will show both the extent and connectivity of these structures between the outgoing CME and post-eruption flaring arcade. Although the upcoming Solar Orbiter mission will make the first inroads towards addressing these problems, its modest maximum latitude of 34° does not result in a significantly different view of coronal structures in regards to ecliptic observations. Besides, those extreme latitudes will last for only a few days making it difficult to study the intermittent solar eruptive activity from that platform. The best science return can be achieved from long-term synoptic observations from high latitudes as envisioned in the SOLARIS white paper proposed in May 2013 to the European Space Agency in response to its call for L2/L3 mission ideas.

## II.   Scientific Objectives

The outstanding issues confronting our current understanding of the solar dynamo and its manifestation in the corona may be summarized through several key scientific questions:

- How is the global, cyclic, solar magnetic field generated?
- What is the nature of flows in the polar regions of the Sun and how do they interact with magnetism?
- Which is the degree of coupling between the magnetic field generated in the northern and southern hemispheres of the Sun?
- How does the radiative energy output of the Sun depend on latitude?
- How does the solar dynamo work and drive connections between the Sun and the heliosphere?
- What is the 3D structure of the global eruptive and quiescent corona?

Progress on these scientific questions requires detailed observations of the solar polar regions, where data is currently scarce and where much of the subtle interplay between plasma flows and magnetic fields that gives rise to cyclic polarity reversals is thought to occur. The out-of-ecliptic observations of the Sun, for the first time, will provide an opportunity for detailed investigations of the magnetic structure and dynamics of the polar region. High-latitude photospheric observations will also provide an unprecedented vantage point for helioseismic imaging that can be used to probe flows and fields in the deep convection zone and tachocline where solar activity is ultimately thought to originate.

In addition to measurements at the photospheric level, the structures of the outer solar atmosphere in polar region and the heliospheric structures merit observations from outside the ecliptic. The poles of sun undergo dramatic change during the 11-year solar cycle, driven by the dynamo action in the solar convection zone. The polar vantage point gives unique opportunities for understanding the origin of the fast solar wind spectroscopically and for stereo viewing of surface vector magnetic fields, coronal structures, and Earth-directed CMEs in coordination with observatories near the Earth. The unique inclination for the SOLARIS observatory will also permit unprecedented measurements of the total solar irradiance. This may help resolve the discrepancy of the cycle variation of the solar irradiance of ~0.1% while solar analogues vary, on average, by 0.3%.

With this in mind, we propose the following prime measurement targets for the SOLARIS mission:

T1) Photospheric magnetic flux distribution and evolution in the polar region

T2) Dynamical coupling between magnetic fields and flows

T3) High-precision measurement of total solar irradiance

T4) Dynamics of the background and transient solar winds

T5) Latitudinal variations of the EUV irradiance and coupling with the magnetic activity

### III. Instrument and Mission Requirements

SOLARIS from its highly inclined orbit around the Sun aims to combine helioseismic and magnetic observations, solar irradiance measurements and EUV images at various latitudes. A total mass of 35 to 50 kg is envisaged for the following three instruments.

The highest priority science objective of SOLARIS is measuring subsurface flow at high-latitude regions, by local helioseismology techniques, based on Dopplergrams acquired by an HMI-MDI type instrument. From this measurement we will derive differential rotation, meridional circulations and convective flows in the upper convection zone. These flow-field measurements will then be cross-correlated with surface magnetic field measurement, to reveal how magnetic flux is transported to the polar region, and how the polarity reversals take place as interplay between plasma flows and magnetic fields. The predominantly vertical kG-field patches that Hinode has found in the polar regions are large enough to be observed. A serious attempt will also be made for stereoscopic helioseismology, for investigating deeper layers, including the best ever shot of the solar tachocline region. Space heritage on this type of instrument is strong if we include also the lighter development done for the Polarimetric and Helioseismic Imager of Solar Orbiter.

Solar irradiance measurements at various latitudes will, for the first time, enable us to measure the anisotropy of the total solar irradiance. The unexplained low photometric variability of the sun may be explained by higher variability at higher latitudes, likely caused by faculae. If it is not the case, then we must conclude that the Sun is a rather atypical star, which will lead to more fundamental questions in astrophysics. Space heritage is also strong given the various irradiance monitor flown on VIRGO, PICARD and other missions.

The wide-field EUV telescope must be able to produce images of the corona in at least two EUV wavelengths (e.g. 30.4 nm and 17.4 nm) and cover a field of view (FOV) of 6 -10 degrees (~10-15 $R_s$ from 0.4 AU) centered on the Sun. Over the past 10 years, the quality of the EUV optics has improved to such a degree that we can seriously contemplate using EUV imagers, that are lighter, more compact and easier to build, to replace low corona coronagraphs whose design remains challenging due to the intrinsically high contrast between the solar disk and the corona at visible wavelengths. Indeed the most recent developments in EUV technologies make it possible to design and build a wide field of view (5 degrees or more), compact (shoebox size) and lightweight (5kg) multi-band EUV telescope. The instrument will be based on the compact design of the Full Sun Imager (FSI) on board Solar Orbiter [3, 6]. FSI includes a movable occulting disk to limit the level of instrumental stray-light for observations above $3R_s$. The 30.4 nm and 17.4 nm bands are dominated by spectral lines formed by resonant scattering above 2 to 3 $R_s$, which produces a signal proportional to the electron density, as in white light coronagraphs. Several technical options, such as using a smaller pitch detector, will be studied to make the design even more compact than on Solar Orbiter

The main mission requirements are as follows:

- Data collection for the primary science goals shall begin within five years of launch
- Data collection for the primary science goals shall begin when the payload can observe above ±60 degrees of solar latitude
- In an orbit passing directly over the solar poles, the pole shall be continuously visible at an observation-zenith angle of less than 30 degrees (TBC) for one full rotation period
- The orbit shall have a very low eccentricity < 0.2 (close to a circular orbit).
- EUV observations will be performed at all latitudes with primary data collection at latitudes ±45 degrees of solar latitude.
- At most two spacecraft for having a 3D view of the corona.

These requirements are much more severe than those put on the Solar Orbiter mission which is not suited for long helioseismic observation and solar irradiance measurements mainly due to the lower inclination to the highly eccentric orbit (>0.8).

# IV. Solar-sail mission design

In the framework of a study led by DLR and ESA, the Gossamer working group (GWG) was set up in order to study how missions based on the use of solar sail could be feasible. The GWG is led by M. Macdonald (Chair and engineer PI) and by T. Appourchaux (Science PI) assisted of several European engineers and scientists comprising L. Gizon (Max Planck Institute for Solar System Research) and T. Sekii (NAOJ). The activities started in November 2011 and will continue beyond the present workshop on Solar Sail.

Table 1: Summary and comparison of solar sail mission architecture options and requirements. Superscript number indicates number of trajectory phases.

|  | Architecture option | | | |
| --- | --- | --- | --- | --- |
|  | $A^2$ | $B^2$ | $C^2$ | $D^2$ |
| Solar pole maximum OZA in one-sidereal rotation period (in deg) | 50 | 40 | 30 | 30 |
| Target solar radius (in AU) | 0.393 | 0.447 | 0.550 | 0.550 |
| Target orbit period (in years) | 0.25 | 0.30 | 0.41 | 0.41 |
| Required sail characteristic acceleration (in mm s$^{-2}$) | 0.2843 | 0.3103 | 0.3655 | 0.5300 |
| Time to target solar radius, i.e. phase 1 duration (in years) | 2.44 | 2.16 | 1.68 | 1.16 |
| Phase 2 duration, i.e. time to reach 60 deg. solar latitude (in years) | 2.56 | 2.84 | 3.32 | 2.29 |
| Time to 60 deg. solar latitude (in years) | 5.00 | 5.00 | 5.00 | 3.45 |
| Time to 90 deg. solar latitude (in years) | 6.75 | 6.94 | 7.25 | 5.00 |

The preliminary mission analysis assumes that the orbit is reached using a two-phase approach: Phase 1 during which a solar sail is deployed and the orbit radius is reduced; Phase 2 during which the orbit is cranked to 90 degrees. Table 1 provides the results for different options. Table 2 provides the spacecraft and payload characteristics only for option $A^2$. With a solar sail of 100 m, a payload mass of 35 kg is enabled within a total launch mass of about 320 kg. Increasing the sail size to 125 m, a payload mass of 60 kg is enabled within a total launch mass of about 500 kg.

Table 2: Summary of mission architecture option $A^2$ mass allowance estimates, including ROM payload allowance estimate.

| Architecture option $A^2$ | | | | | | |
| --- | --- | --- | --- | --- | --- | --- |
| Sail Side Length (m) | S/C Mass Range with 7.5 µm film (kg) | S/C Mass Range with 2.5 µm film (kg) | Platform Mass range with 7.5 µm film (kg) | Platform Mass range with 2.5 µm film (kg) | Payload Mass with 7.5 µm film (kg) | Payload Mass with 2.5 µm film (kg) |
| 50 | *120 – 107* | *95 – 83* | *negative* | *negative* | *negative* | *negative* |
| 75 | *202 – 172* | 146 – 116 | < 13 | 39 – 69 | < 5 | 10 – 15 |
| 100 | 306 – 255 | 207 – 156 | 23 – 73 | 122 – 172 | 5 – 15 | 25 – 35 |
| 125 | 435 – 360 | 281 – 205 | 78 – 154 | 232 – 308 | 15 – 30 | 45 – 60 |

## V. Interest of the scientific community

The French community interested in this project is located in the Paris area (Institut d'Astrophysique Spatiale, Orsay; Laboratoire d'Etudes Spatiales et d'Instrumentation en Astrophysique, Meudon; Institut de Recherche sur les lois Fondamentales de l'Univers, Saclay). These laboratories are involved in the Solar Orbiter mission and were involved in the study of the Plan A of Solar-C. Additional laboratories that are interested are the Observatoire de la Côte d'Azur and Institut de Recherche en Astrophysique et Planétologie in Toulouse.

The international community interested is the same as the one that proposed several out-of-ecliptic concepts such as Solar Polar Investigation (SPI), POLARIS or the Plan-A of Solar-C. These are namely the Jet Propulsion Laboratory (USA), Stanford University (USA), the Naval Research Laboratory (NRL), the Max-Planck institute for Solar System Research, the World Radiation Center (Switzerland) and the NAOJ (Japan), the INAF-Astrophysical Observatory of Turin (Italy). China is also potentially interested since they are considering the SPORT mission that is also an out-of-ecliptic mission.

## VI. Conclusion

The POLARIS mission was submitted in 2007 as L-class mission to the Cosmic Vision programme of ESA [2]. It was rejected based on the low technology readiness of the solar sails. The Gossamer road map established by the Working Groups of DLR/ESA aims at the successful deployment of a solar sail system of 50 m square by 2018 (Gossamer-3). Therefore we can reasonably envisage a demonstrated solar-sail technology by 2020 for a potential mission in the framework 2025-2030. The *Solaris* concept developed in this paper was submitted both to the Centre National d'Etudes Spatiales in January 2013 and to the European Space Agency in May 2013. Such a mission would require a large international collaboration involving several agencies (ESA, JAXA, NASA, China). A mission of the L-class allowing a payload of about 50 kg with solar sails of 100 m to 150 m seems realistic. Nevertheless, at the time of writing, there is not yet any programmatic window either in Europe or in the US, but this could change in the Fall of 2013 if a mission using solar sail is selected by ESA.

The prospect of having a potential dual-spacecraft solar mission out of ecliptic using solar sail is an exciting opportunity for great scientific return with large discovery-like potential. This opportunity and challenge shall not be missed.

## Acknowledgments

TA gratefully acknowledges support by the Centre National d'Etudes Spatiales. DM and AV are supported by NASA and ONR funds.